\documentclass{article}
\usepackage[T1]{fontenc} 
\usepackage[utf8]{inputenc} 
\usepackage{ismir,amsmath,cite,url}
\usepackage{graphicx}
\usepackage{color}
\usepackage{pgfplots}
\usepackage{amsmath}
\usepackage{enumitem}
\usepackage{amsfonts}
\usepackage{multirow}
\usepackage{xcolor}

\title{Exploiting Pre-trained Feature Networks for Generative Adversarial Networks in Audio-domain Loop Generation}

\multauthor
{Yen-Tung Yeh$^{1,2}$  \hspace{1cm} Bo-Yu Chen$^{1,2}$ \hspace{1cm} Yi-Hsuan Yang$^{1, 3}$} 
{\\
$^1$~Academia Sinica, 
$^2$~National Taiwan University,  
$^3$~Taiwan AI Labs \\
{\tt\small ytsrt66589@gmail.com, bernie40916@gmail.com, yang@citi.sinica.edu.tw}
}  

\def\authorname{Y.-T. Yeh, B.-Y. Chen, and Y.-H. Yang}
\usepackage[bookmarks=false,pdfauthor={\authorname},pdfsubject={\papersubject},hidelinks]{hyperref}

\begin{document}

\maketitle
\begin{abstract}
While generative adversarial networks (GANs) have been widely used in research on audio generation, the training of a GAN model is known to be unstable, time consuming, and data inefficient. Among the attempts to ameliorate the training process of GANs, the idea of Projected GAN emerges as an effective solution for GAN-based image generation, establishing the state-of-the-art in different image applications. The core idea is to use a pre-trained classifier to constrain the feature space of the discriminator to stabilize and improve GAN training. This paper investigates whether Projected GAN can similarly improve audio generation, by evaluating the performance of a StyleGAN2-based audio-domain loop generation model with and without using a pre-trained feature space in the discriminator. Moreover, we compare the performance of using a general versus domain-specific classifier as the pre-trained audio classifier. With experiments on unconditional one-bar drum loop and synth loop generation, we show that a general audio classifier works better, and that with Projected GAN our loop generation models can converge around 5 times faster without performance degradation.
\end{abstract}

\section{Introduction}\label{sec:intro}

Generative adversarial networks (GANs) \cite{goodfellow2014gan} are deep generative models that are composed of a \emph{generator} and a \emph{discriminator}. The discriminator can be considered as a classifier (or a ``critic'') which judges whether its input is a real sample, or a synthetic one created by the generator counterpart; the generator takes as input a random vector (sometimes plus additional conditional inputs \cite{mirza2014}) and aims to create a synthetic sample that ``fools'' (i.e., appears to be realistic to) the discriminator. During the GAN training process, the discriminator and generator are updated in an iterative manner, fixing the parameters of one and updating those of the other each time. After the training converges, the generator can be used to generate original samples, leading to state-of-the-art (SOTA) results in image generation \cite{biggan,stylegan,stylegan2,stylegan3,styleganxl} and many other domains. 

GANs have been extensively applied to music generation as well, including symbolic-domain generation \cite{crnngan,yang2017midinet,dong2018musegan,chen18mm,jazzgan18,wei19ismir,jhamtani19,9132664,yuyi21tomccap,transformergan} 
and audio-domain generation \cite{engel2019gansynth,wavegan19iclr,wgansing,lattner2019highlevel,drysdale2020adversarial,unagan,drumgan,nistal2021vqcpcgan,darkgan21ismir,hung21ismir,djtransgan}.
In particular, GAN-based models represent the SOTA in audio-domain music generation tasks such as single-note generation \cite{engel2019gansynth,darkgan21ismir}, drum track generation \cite{lattner2019highlevel}, and loop generation \cite{hung21ismir}.


Despite its widespread applications, GANs are notoriously difficult to train \cite{salimans2016,arjovsky2017principle,kodali2017,gulrajani2017improved,mescheder2018}. 
This is partly due to the fact that the generator and discriminator have opposite goals by design, with the generator aiming to \emph{maximize} the discriminator loss and the discriminator aiming to \emph{minimize} the same loss in different iterations, making the training dynamics complicated. 
As the parameters of the discriminator are constantly being updated over the training process, 
the generator has no single critic to improve its own performance over time.
And, while the parameters of the generator and discriminator are typically initialized \emph{randomly},
the GAN training process can be time-consuming.
Using a pre-trained feature network as part of the discriminator has been shown to speed up the training process 
\cite{Zhaoicml20, grigoryeviclr22, Kumaricvpr22}, 
but some modifications of the pre-trained feature network is  needed to avoid the discriminator from being too strong and causing the gradients of the generator to vanish.

Projected GAN \cite{Sauer2021NEURIPS} is a new approach that is shown to effectively leverage the benefits of pre-trained feature networks, leading to SOTA results in unconditional image generation \cite{styleganxl}. Projected GAN adds two \emph{random projection} modules 
after the pre-trained feature networks, named ``cross-channel mixing'' (CCM) and ``cross-scale mixing'' (CSM), to prevent the discriminator from focusing only on a subset of features and thereby avoid mode collapse. 
They not only reduce the time of GAN training, but also improve the quality of the generated images.


This paper studies whether Projected GAN can also improve audio generation, by incorporating it to unconditional audio-domain loop generation  \cite{hung21ismir}  as a case study.
We note that there are well-studied pre-trained feature networks in the image domain \cite{Zhaoicml20, grigoryeviclr22, Kumaricvpr22}.
There are also studies 
on the choice of pre-trained network for retrieval and analysis of musical audio \cite{Choi17ismir} and soundtracks \cite{45611}.
However, we are less sure which pre-trained feature networks to use in the context of musical audio generation.
Therefore, besides pioneering the use of Projected GAN for audio generation, an interesting aspect of our research is that we investigate different types of pre-trained feature networks for Projected GAN-based musical audio generation, including general classifier and domain-specific classifiers.  


Specifically, 
for the \textbf{general} classifier, 
we use the pre-trained VGGish feature networks trained on YouTube-100M \cite{45611}, a huge collection of sounds and musical audio from YouTube. 
For \textbf{domain-specific} classifiers, we use short-chuck convolutional neural network (SCNN)-based models \cite{won2020evaluation}, which have led to SOTA accuracy across multiple music auto-tagging datasets. 
We use two SCNNs, one trained on the MagnaTagATune (MTAT) dataset \cite{9aed49b956a24e99b044582665fd5b21} for tagging music of a wide range of genres, and the other trained on a collection of loops for genre classification of loops, which are domain-wise even closer to our generation task. 
We elaborate on the datasets and  classifiers in Section \ref{sec:db}, and then the usage of the classifiers as the pre-trained feature networks for loop generation in Section \ref{sec:model}.

We report objective and subjective evaluations in Sections \ref{sec:exp} and \ref{sec:exp2} studying the influence of different pre-trained feature networks for drum loop generation and synth loop generation following the approach of Projected GAN, finding that the use of a general classifier works consistently better.
Moreover, we find slightly
better result can be obtained if we ``fuse'' general and domain-specific pre-trained feature networks in our model. 
Objective and subjective evaluations both demonstrate that Projected GAN improves training speed and the final generation quality.

\begin{figure}[t]
\centering
\includegraphics[width=.97\columnwidth]{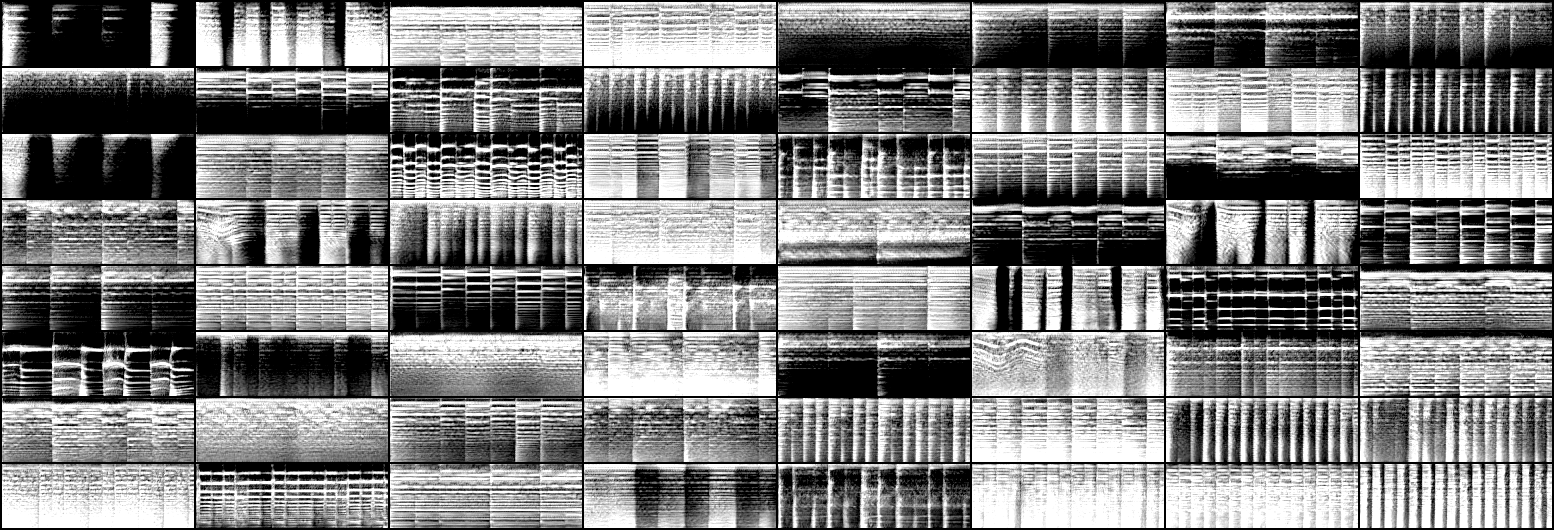}
\vspace{-2mm}
\caption{Mel-spectrograms of examples of synth loops generated by Projected GAN with a general classifier.}
\label{fig:synth_melspec}
\end{figure}

We share our code at 
\url{https://github.com/Arthurddd/pjloop-gan},
and examples of the generated loops at 
\url{https://arthurddd.github.io/PjLoopGAN/}.
Illustrations of the Mel-spectrograms of the generated synth loops and drum loops can be found respectively in Figure \ref{fig:synth_melspec} and later on in Figure \ref{fig:melspec}. 

\section{Background}\label{sec:bg}

\noindent
\textbf{Related Work.}
GANs have garnered great interest in recent years, leading to models that generate 
high-fidelity audio waveforms. 
WaveGAN \cite{wavegan19iclr} pioneers the use of GAN for audio; it can synthesize one-second raw audio waveforms of speech and music with coherence.
Follow-up research applies GAN to other audio synthesis tasks. For example, MelGAN \cite{melgan} and Parallel WaveGAN \cite{pwg} 
use GAN to build neural vocoders that can reconstruct a waveform from the corresponding Mel-spectrogram. 
GAN-TTS \cite{gantts} applies 
GAN to convert text to natural human speech. UNAGAN \cite{unagan} uses GAN to synthesize  singing voice.

GAN has also been used to synthesize audio samples of musical materials, including percussive and harmony ones, that can be used in music production. 
GANSynth \cite{engel2019gansynth} can synthesize harmony music notes with diverse timbre and controllable pitches.
DrumGAN \cite{drumgan} 
can synthesize one-shot percussion sounds, allowing for the use of condition perceptual features to intuitively control the timbre of the percussion sounds. 
StyleWaveGAN \cite{stylewavegan} improves the audio quality and inference time of DrumGAN.
However, the aforementioned models deal with only single notes or one-shot samples instead of longer phrases such as musical loops, limiting their applicability to creating loop-based music. In view of this need, Hung \emph{et al.} \cite{hung21ismir}
study the task of one-bar drum loop generation and benchmark the performance of UNAGAN \cite{unagan}, StyleGAN \cite{stylegan} and StyleGAN2 \cite{stylegan2} for this task over the FreeSound Loop dataset \cite{ramires2020freesound}, showing that the model based on StyleGAN2 \cite{stylegan2} performs the best. 
However, Hung \emph{et al.} \cite{hung21ismir} do not consider the training efficiency of their models.
Moreover, while they focus on drum loops only, we consider also the generation of synth loops to encompass not only percussive but also harmonic musical materials in our work.

Specific to the GAN training technique, 
there are three main approaches to improve the discriminator. First, modifying the \emph{architecture} of the discriminator to improve the discriminator's ability \cite{ebgan,unetgan,transgan}.
Second, assembling \emph{multiple discriminators} to capture more comprehensive features, an approach that as been widely investigated in the audio domain for building the vocoder \cite{melgan,vocgan,hifigan,fregan}. 
The third approach introduces a \emph{pre-trained feature network} to the discriminator to avoid learning its parameters completely from scratch, which helps speed up the convergence time and prevent model overfitting. 
Zhao \emph{et al.} \cite{Zhaoicml20} use the pre-trained network in the large-scale dataset, adapt it to a small dataset, and propose adaptive filter modulation to deal with domain shift. Grigoryev \emph{et al.} \cite{grigoryeviclr22} and Kumari \emph{et al.} \cite{Kumaricvpr22} both find that pre-trained network selection can largely influence performance and propose a recipe to choose a suitable pre-trained network for a particular image-domain task. 
Projected GAN \cite{Sauer2021NEURIPS} is a very recent idea that combines the aforementioned approaches, using multiple discriminators and employing pre-trained feature networks simultaneously to improve the training stability and efficiency of the GAN. 
However, to our best knowledge, adapting Projected GAN to the audio-domain generation is still unexplored, for either speech or music. 

\vspace{5mm}
\noindent
\textbf{Projected GAN.}
Projected GAN 
\cite{Sauer2021NEURIPS}
introduces a set of $L=4$ feature projectors $\{P_l, l=1,\dots,L\}$, each maps either real samples $\mathbf{x}$ or fake samples $G(\mathbf{z})$ to a fixed pre-trained feature space. It aims to minimize the following objective function to match the data distribution in the feature space, instead of matching data distributions directly:
\begin{equation}
\begin{split}
 \mathop{\min}_{G}\mathop{\max}_{\{D_l\}}\sum\nolimits_{l=1}^L \exp\big(E_x[\log D_l(P_l(\mathbf{x}))] \\ +E_z[\log(1-D_l(P_l(G(\mathbf{z}))))] \big) \,,
 \end{split}
\end{equation}
where $\{D_l, l=1,\dots,L\}$ denotes the set of independent discriminators operating on different feature projections, and $G$ denotes the generator that converts a random vector $\mathbf{z}$ into a fake sample. 
Each projector $P_l$ consists of three components: a pre-trained feature network $C_l$, the CCM modules, and the CSM modules. 
First, we extract features from $L$ different layers of  the feature network $C_l$, leading to $L$ set of feature maps in different scales. Second, for each scale, the CCM mixes the features across channels by \emph{random} (i.e., not-learned) $1 \times 1$ convolutions with an equal number of input and output channels, which can be viewed as the generalization of random permutation.
Third, CSM further mixes features across scales by \emph{random} $3 \times 3$ convolutions and bilinear upsampling layers, yielding a U-Net architecture that combines the feature maps of different scales. The features fused by CCM and CSM in $L$ different scales would then be fed to each discriminator $D_l$.
The random projections introduced by CCM and CSM have the effect of encouraging each $D_l$ to take into account the entire feature space instead of overfitting to a sub-set of feature space, thereby avoiding mode collapse. 
Without using gradient penalties and any sophisticated training strategies, Projected GAN updates its loss simply by summing the output of the $L$ discriminators. 


\section{Datsets \& Classifiers}\label{sec:db}

Our work is built upon the use of the following datasets.

\vspace{1mm}
\noindent\textbf{Youtube-100M} \cite{45611} is a private dataset of Google, containing 100 million Youtube videos.  
Each video has on average 5 manually assigned tags, out of 30,871 possible labels.
Hershey \emph{et al.} \cite{45611} train a VGGish pre-trained network for large-scale audio classification using the dataset. 
While we are not able to have a copy of the dataset, we can use the pre-trained weights Gemmeke \emph{et al.}\cite{audioset} share publicly as our general audio classifier.

\vspace{1mm}
\noindent\textbf{MagnaTagATune (MTAT)} \cite{9aed49b956a24e99b044582665fd5b21} is a dataset commonly-used in research on music auto-tagging.  It contains 25,863 music clips, each 29-seconds long. 
We use MTAT to train one of our domain-specific pre-trained feature networks.
We follow the original data split \cite{9aed49b956a24e99b044582665fd5b21} and  use only the top 50 tags, including genre and instrumentation labels, as well as decades (e.g., `80s' and `90s') and moods. 

\vspace{1mm}
\noindent\textbf{Looperman dataset} is an in-house collection of loops from \url{https://www.looperman.com/}, a website hosting free music loops.
We get the audio and uploader-provided metadata of 23,983 drum loops and 22,625 synth loops. 
According to the metadata, the drum loops and synth loops can be categorized to 66 and 58 genres, respectively.
We use the Looperman dataset for not only building a  domain-specific pre-trained feature network but also for building our audio-domain loop generation model.

Following the preprocessing steps of  Hung \emph{et al.} \cite{hung21ismir}, we first apply downbeat tracking via \texttt{madmom} \cite{madmom} to split every loop into 
single bars and then time-stretch each  to 120 BPM (beats-per-minute) by \texttt{pyrubberband} \cite{pyrubberband} to unify their length to be always \textbf{2 seconds per loop}. 
After this preprocessing, we have 128,122  and 42,570 one-bar loops for drum and synth, respectively. We split the data by 80/10/10 for training the loop genre classifier, and use the entire data for music loop generation.



\begin{figure}[t]
\centering
\includegraphics[width=.86\columnwidth]{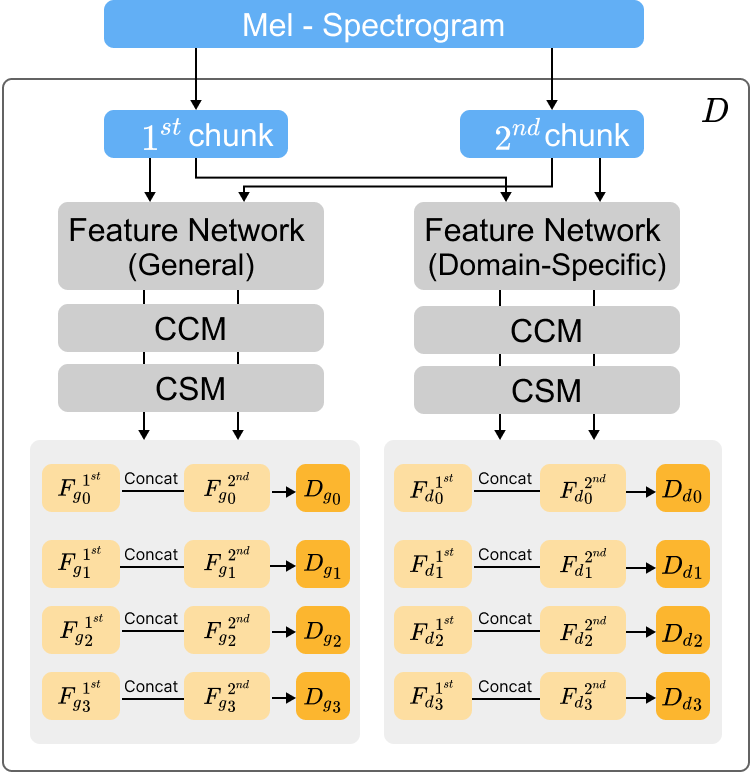}
\vspace{-3mm}
\caption{Diagram of the proposed discriminator architecture. A Mel-spectrogram is divided into two chunks and then fed to either a general or a domain-specific pre-trained feature network, or both (for ``fusion'').  Features computed for both chunks are aggregated at each scale $l$ before feeding to the corresponding discriminator $D_l$. }
\label{fig:model}
\end{figure}

\subsection{Pre-trained Feature Networks for Music}
\label{subsec:featurenetworks}


\textbf{General.} As aforementioned, we use the VGGish network trained on Youtube-100M 
\cite{audioset} as the general pre-trained feature network. Taking the Mel-spectrogram as input, the model uses 2D convolutional blocks and 2D max-pooling layers to compute 128-dimensional features at the output. 

\vspace{1mm}
\noindent\textbf{Domain specific.} Short-Chuck CNN (SCNN)\cite{won2020evaluation} has a simple 2D CNN architecture with $3\times 3$ filters and residual module but it has been shown to be remarkably effective for music auto-tagging. We use in our implementation  6 layers of CNN blocks with a fully connected layer and the residual module. Each CNN block comprises a $2\times 2$ max-pooling layer.  
We use SCNN to train separate classifiers for MTAT and Looperman  from scratch, and then used the trained classifiers as our domain-specific pre-trained feature networks.
For MTAT, SCNN achieves 0.909 ROC-AUC and 0.445 PR-AUC.
For genre classification of the looperman drum loops, 
the classification accuracy reaches 0.792.
For synth loops, the accuracy attains 0.700.


\vspace{1mm}
\noindent\textbf{Fusion.} 
Fusing multiple pre-trained feature networks has also been shown useful for Projected GAN-based image generation \cite{styleganxl}.
Accordingly,
we also experiment with a simple fusion strategy that allows the discriminator to consider both general and domain-specific features at the same time, as depicted in Figure \ref{fig:model}. 

\section{Audio Generation by Projected GAN}\label{sec:model}

Following Hung \emph{et al.} \cite{hung21ismir}, we use  StyleGAN2 \cite{stylegan2} as the backbone of our loop generation model. However, we improve their model in a number of aspects, regarding to not only the discriminator (i.e., with Projected GAN) but also the generator. We provide the details below. 

\begin{figure}[t]
\centering
\includegraphics[width=.95\columnwidth]{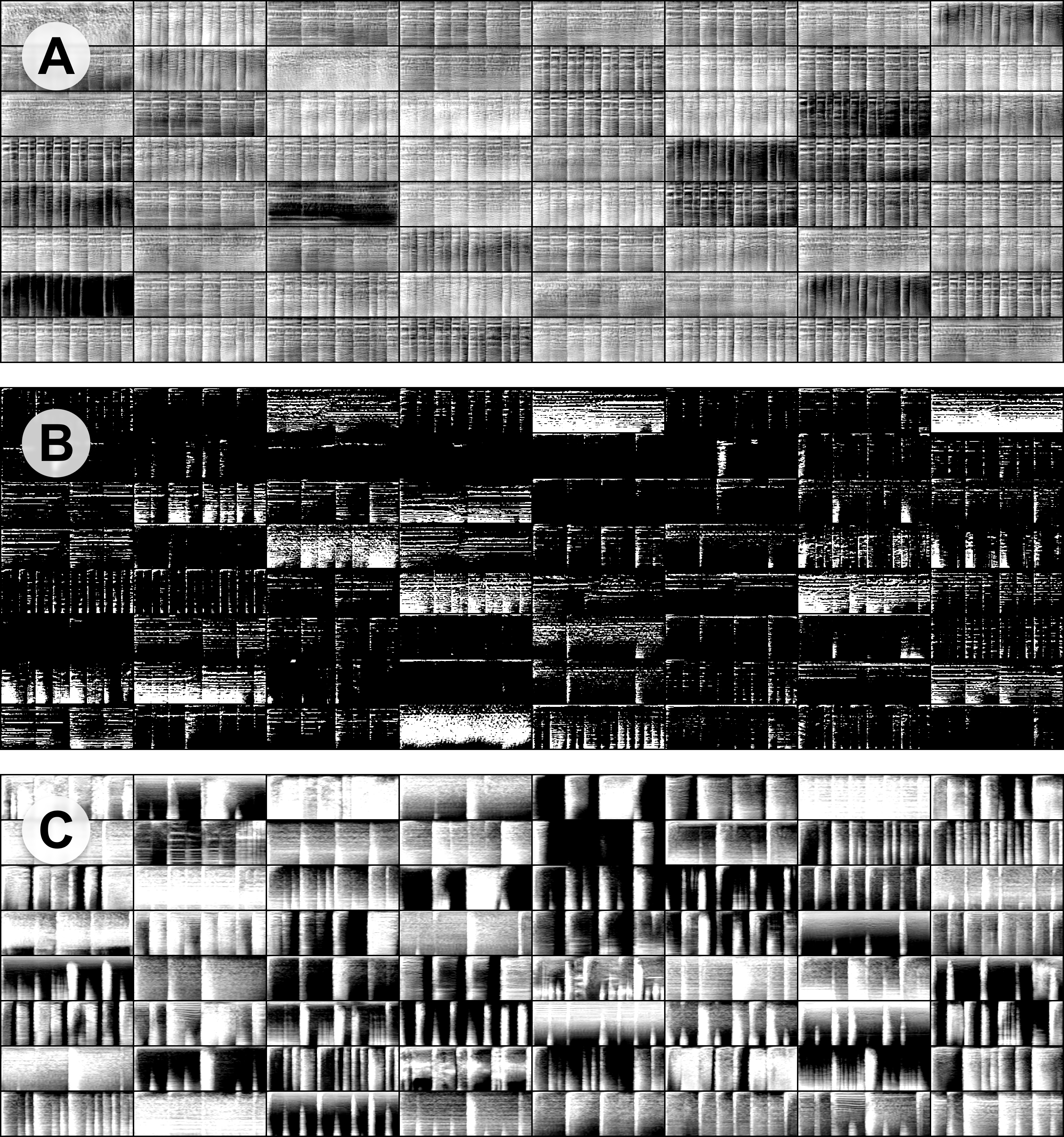}
\vspace{-2mm}
\caption{Mel-spectrograms of examples of drum loops generated (a) by
a failed case using an overly large latent space for $\mathbf{z}$, leading to mode collapse (see Section \ref{sec:model}); (b) Projected GAN with the MTAT domain-specific classifier; (c) Projected GAN with the VGG general classifier.}
\label{fig:melspec}
\end{figure}

\vspace{1mm}
\noindent\textbf{Improving the generator.}
The generator $G$ consists of a mapping network $G_m$ and a synthesis network $G_s$. We implement $G_m$ with only 6 fully-connected layers instead of the 8 in the original StyleGAN2 architecture. Furthermore, echoing the finding in the image domain \cite{DBLP:journals/corr/abs-2104-08894}, 
we find that the length of the vectors $\mathbf{z}$ (i.e., the input of $G_m$) largely affects the model performance. 
Setting the length of $\mathbf{z}$ to 512 as done in \cite{hung21ismir} leads to mode collapse in our preliminary experiments, as shown   
in Figure \ref{fig:melspec}(a). 
This may be related to the so-called ``intrinsic dimension'' of the data \cite{DBLP:journals/corr/abs-2104-08894}; an overly large latent space of $\mathbf{z}$ introduces redundancy that distracts the generator. 
As smaller $\mathbf{z}$ empirically works better, we set its length to 32 in the experiments reported below. 
We similarly set the length of the ``style code'' $\mathbf{w}$ (i.e., the output of $G_m$) to a small value of 64. 


\vspace{1mm}
\noindent\textbf{Pipeline}
The training follows the procedure of the Projected GAN, but we use the following pipeline to match the input shape expected by the pre-trained feature networks.
First, we compute the Mel-spectrogram with 512-point window size and 160-point hop size for short-time Fourier Transform (STFT) and 64 Mel channels. 
Second, we feed a latent $\mathbf{z}$ to the mapping network $G_m$ to generate style codes that modulate the convolutions of the synthesis network $G_s$, using four upsampling blocks to generate a Mel-spectrogram that corresponds to a synthesized \emph{2-second} loop. 
We note that our pre-trained feature networks are on \emph{1-second} audio.
To address the size mismatch, 
we split the Mel-spectrogram into \textbf{two chunks}, 
with the first half corresponding to the first second and the other the rest. 
As illustrated in Figure \ref{fig:model}, we feed the two chunks separately to the discriminators, going through the pre-trained feature network, CCM and CSM.
Moreover, we concatenate the features with the same scales but in different chunks to aggregate the features from individual chunks, yielding $L=4$ aggregated features $\{F_l, l=1...L\}$. We feed each of these features independently to the corresponding discriminators $D_l$. Additionally, if we consider two pre-trained feature networks simultaneously, referred to as ``fusion'' in Section \ref{subsec:featurenetworks}, we have in total $2L$ aggregated features and $2L$ discriminators.
Eventually, we sum the loss from these discriminators to update the network.

Similar to Hung \emph{et al.} \cite{hung21ismir}, at inference time, the Mel-spectrogram generated by the generator would go through a MelGAN vocoder \cite{melgan} to become waveforms.


\begin{table*}
\centering
\scalebox{1}{
\begin{tabular}{ |l | r r r r | r r r r |} 
 \hline
 \multirow{2}{*}{\textbf{Models}} & \multicolumn{4}{|c|}{\textbf{Drum loops}} & \multicolumn{4}{|c|}{\textbf{Synth loops}} \\
  & {IS}~$\uparrow$  & {FAD}~$\downarrow$ & {D}~$\uparrow$ & {C}~$\uparrow$ & {IS}~$\uparrow$  & {FAD}~$\downarrow$ & {D}~$\uparrow$ & {C}~$\uparrow$ \\ 
 \hline
 {\scriptsize{A}~} Real data &16.30  & 0.01  & 1.00 & 1.00 &14.95  & 0.01  & 1.00 & 1.00 \\ 
 \hline
 {\scriptsize{B}~} StyleGAN2 \cite{hung21ismir} & 5.58  & 2.50  & 1.01 & \textbf{0.89} & 4.80  & 3.60  & 1.19 & \textbf{0.91}\\
 {\scriptsize{C}~} StyleGAN2 (early-stop) & 5.21  &  3.40 &  0.56 & 0.43  &  4.55 & 7.97 & 1.01 & 0.75 \\
 \hline
 {\scriptsize{D}~} Projected StyleGAN2 (VGG) & 5.87  & 3.03  & 1.06 & 0.70 & 4.70  & \textbf{2.34}  & 1.31 & 0.82\\ 
 {\scriptsize{E}~} Projected StyleGAN2 (SCNN$_\text{MTAT}$) & 4.75  & 8.18 & 0.00 & 0.00 & 3.97  & 7.14 & 0.001 & 0.002 \\ 
 {\scriptsize{F}~} Projected StyleGAN2 (SCNN$_\text{Loop}$) & 4.45  & 7.21  & 0.00 & 0.00 & 3.08  & 8.32  & 0.001 & 0.002 \\
 \hline
 {\scriptsize{G}~} Projected StyleGAN2 (VGG$+$SCNN$_\text{MTAT}$)  & 6.22  & 2.45  & \textbf{1.11}  & 0.74 & 4.73  & 3.13  & 1.67  & 0.85 \\
{\scriptsize{H}~} Projected StyleGAN2 (VGG$+$SCNN$_\text{Loop}$)  & \textbf{6.31}  & \textbf{2.34}  & 1.08  & 0.73 & \textbf{4.82}  & 2.56  & \textbf{1.70}  & 0.85 \\
 \hline
\end{tabular}}
\caption{Objective evaluation result for the different settings of the Projected GANs trained on the Looperman dataset for loop generation. `D' and `C' stand for Density \& Coverage \cite{ferjad2020icml} ($\downarrow$\,/\,$\uparrow$: the lower/higher the better; the best in bold).}
\label{tab:objective}
\end{table*}

\section{Objective Evaluation}
\label{sec:exp}
We use the following objective metrics to evaluate the quality and diversity of the generated loops.

\vspace{1mm}
\noindent\textbf{Inception Score (IS)} \cite{DBLP:conf/nips/SalimansGZCRCC16,barratt2018note}
measures the quality of the generated loops and detects whether there is a mode collapse by using a pre-trained domain-specific classifier, namely the loop genre classifier for each type of loops (i.e., drum or synth). It penalizes models whose samples cannot be reliably classified into a single class or that only belong to a few from all possible classes. 

\vspace{1mm}
\noindent\textbf{Fr\'{e}chet Audio Distance (FAD)} \cite{kilgour2019frechet} 
reflects both quality and diversity as it measures the distance between continuous multivariate Gaussians
fitted to the embeddings of the real and generated loops. 

\vspace{1mm}
\noindent\textbf{Density\,\&\,Coverage~(D\&C)}  \cite{ferjad2020icml} are new metrics that measure respectively the quality and diversity of the generated data.
D\&C are considered to be more robust against the influence of outliers compared to older metrics such as precision and recall (P\&R) \cite{kynkaanniemi2019improved}.
`D' is the average number of real-sample neighborhood spheres that contain each of the fake samples. It may be greater than $1$ depending on the density of reals around the fakes. `C' is the fraction of real samples whose neighborhoods contain at least one fake sample.
In our implementation, we use the hyperparameters suggested by \cite{ferjad2020icml}. Moreover, following \cite{ferjad2020icml}, we calculate D\&C with a randomly-initialized VGG16 model that projects samples to VGG16 \emph{fc2} space whose dimension is deliberately set to a small value 64.


We evaluate the following models here:
\begin{itemize}[leftmargin=*,itemsep=0pt,topsep=2pt]
\item \textbf{StyleGAN2}: the SOTA for drum loop generation \cite{hung21ismir}.
\item \textbf{StyleGAN2~(early-stop)}: the StyleGAN2 that stops training at the same point when Projected GAN converges, providing a reference demonstrating the possible advantage of the training efficiency of  Projected GAN.
\item \textbf{Projected~StyleGAN2} with different pre-trained feature networks, including  1) the general VGG classifier alone, 2) the domain-specific SCNN classifier trained on MTAT alone (denoted as SCNN$_\text{MTAT}$), 3) the SCNN classifier trained on Looperman alone, using the drum or synth classifier depending on whether it is for generating drum or synth loops  (denoted as SCNN$_\text{Loop}$), 4) ``fusion'' of VGG, SCNN$_\text{MTAT}$, and 5) ``fusion'' of VGG, SCNN$_\text{Loop}$.

\end{itemize}

Table \ref{tab:objective} presents the objective evaluation results of models trained on drum loops and synth loops. Each model generates 10,000 random loops to compute the scores. We also compute the scores using the real data for setting a high anchor of the objective scores. 

We see that the best-performing configurations of Projected StyleGAN2 can achieve higher IS and lower FAD than StyleGAN2 (i.e., row B).  For drum loops, the IS can be improved from 5.58 to 6.31 (row H) ; for synth loop the FAD can be reduced from 3.60 to 2.34 (row D).

Interestingly and somehow surprisingly, when only a single pre-trained feature network is used (i.e., rows D--F), we find that \emph{only} Projected StyleGAN2\,(VGG) performs nicely.
The domain-specific pre-trained feature networks actually degrade the performance of loop generation; either Projected StyleGAN2\,(SCNN$_\text{MTAT}$) or Projected StyleGAN2\,(SCNN$_\text{Loop}$) 
obtains lower IS and higher FAD, and even close-to-zero D\&C. 
This  suggests that a general pre-trained feature network works much better than a domain-specific pre-trained feature network in the context of Projected GAN-based unconditional loop generation.
This is likely because the discriminator employing a domain-specific pre-trained feature network is too strong compared to the generator, leading to gradient vanishing. 
Figure \ref{fig:melspec}(b) shows samples generated by Projected StyleGAN2 (SCNN$_\text{MTAT}$); we see that the model weirdly generates sparse samples most of the time.

Table \ref{tab:objective} also shows that the ``fusion'' of general and domain-specific pre-trained feature networks (rows G\,\&\,H) performs slightly better in some metrics than using a general pre-trained feature network alone (row D).
In particular, we see that Projected StyleGAN2 (VGG$+$SCNN$_\text{Loop}$) achieves the highest IS score in both drum and synth loop generation. It also reaches the lowest FAD for drum loops and the second lowest FAD for synth loops. 
We show examples of its generated synth and drum loops in Figures \ref{fig:synth_melspec} and \ref{fig:melspec}(c), respectively.

Projected StyleGAN2 appears to have lower D\&C compared to StyleGAN2, suggesting that Projected StyleGAN2 sacrifices a little diversity for higher quality of the generated samples. 
This is presumably because the feature space has been constrained by the pre-trained feature networks throughout the whole training process, making it hard to capture the whole distributions. Future work can be done to remedy this and further improve the diversity of Projected StyleGAN2.

\begin{figure}[t]
\centering
\includegraphics[width=1\columnwidth]{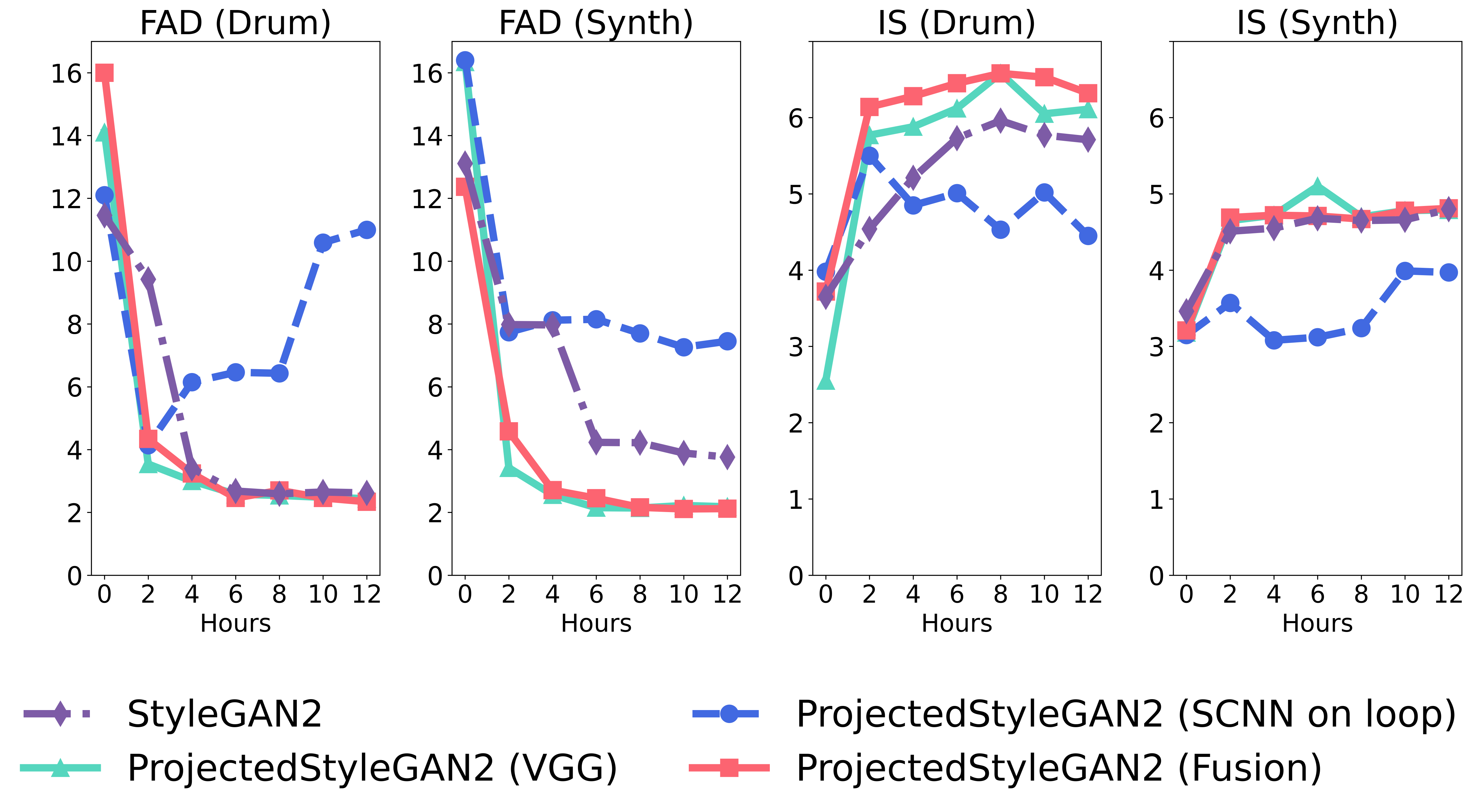}
\vspace{-4mm}
\caption{The FAD and IS as a function of training hours.} 
\label{fig:curve}
\end{figure}

Figure \ref{fig:curve} shows how the FAD and IS of four selected models (rows B, D, F, H in Table \ref{tab:objective}) varies as a function of training time, when all the models are trained separately and independently on a single V100 GPU. For both drum and synth loop generation, Projected GAN leads to lower FAD much faster than the StyleGAN2 baseline. For synth loops, Projected GAN with only 2-hour training reaches lower FAD than the StyleGAN2 baseline with 12-hour training. 
In general, Projected GAN converges at approximately 2 hours for both drum and synth loops, while StyleGAN2 converges x5 times longer at about 10 hours. 
Echoing the result in Table \ref{tab:objective}, using the epoch of StyleGAN2 corresponding to 2-hour training time (i.e., row C) obtains worse scores than the models corresponding to rows B and D  in almost all the metrics for both drum and synth loops. Overall, these results nicely demonstrate how Projected GAN speeds up and improves GAN training.

\begin{figure*}[t]
\centering
\includegraphics[width=1\textwidth]{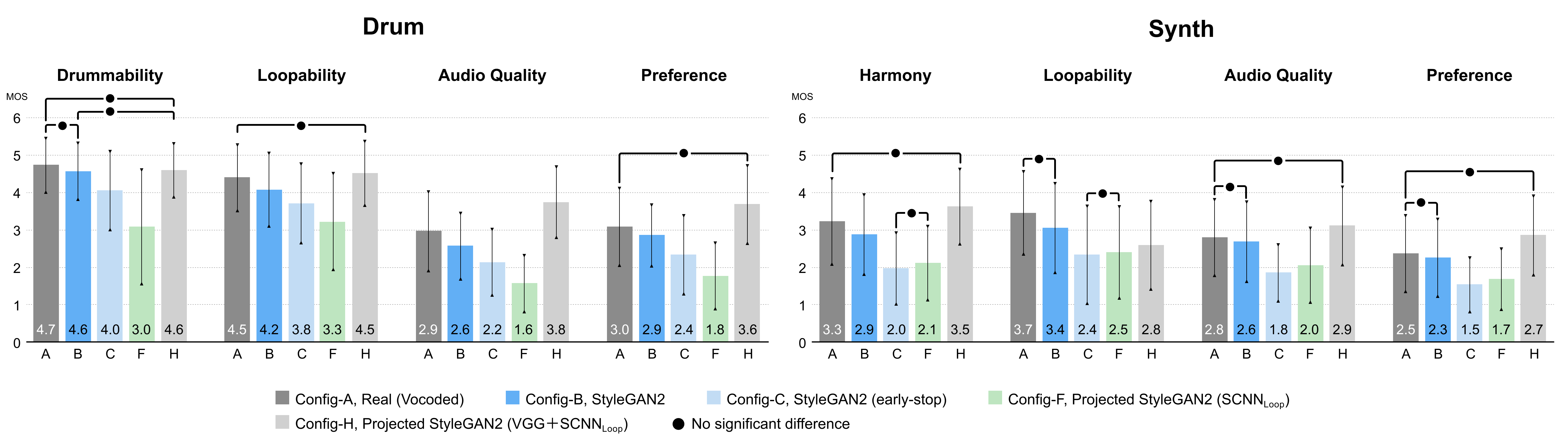}
\vspace{-2mm}
\caption{Subjective evaluation results. The performance difference between any pair of models in any metric is statistically significant ($p$-value$<0.001$) under the Wilcoxon signed-rank test,
except for the pairs that are explicitly highlighted. }
\label{fig:subjective}
\end{figure*}

\section{Subjective Evaluation}
\label{sec:exp2}

To further assess the loops generated by the models, we conduct a subjective listening test through an anonymous online questionnaire. Each subject is presented with a randomly picked human-made loop from the Looperman dataset (i.e., `Real') and a randomly-generated loop by each of the following four models: StyleGAN2, StyleGAN2 (early-stop), Projected StyleGAN2 (SCNN$_\text{Loop}$) and Projected StyleGAN2 (VGG$+$SCNN$_\text{Loop}$). The subject is then asked to rate each of these loops in terms of the following metrics, all on a five-point Likert scale.
\begin{itemize}[leftmargin=*,itemsep=0pt,topsep=2pt]
\item \textbf{Drummability or Harmony}: Drummability (for drum loops) means whether the sample contains percussion sounds, and Harmony (for synth loops) means whether the sample contains harmonic sounds.
\item \textbf{Loopability}: whether the sample can be played repeatedly in a seamless manner.
\item \textbf{Audio quality}: whether the sample is free of unpleasant noises or artifacts.
\item \textbf{Preference}: how much you like it.
\end{itemize}
To evaluate loopability, we repeat each sample four times in the audio recording. 
Because the output of the models all goes through the MelGAN vocoder \cite{melgan} to become waveforms,
we do the same for `Real' to be fair. 

Figure \ref{fig:subjective} shows the averaged results from 35 participants. The responses indicate an acceptable level of reliability, Cronbach’s $\alpha=0.721$. The subjective evaluation result aligns nicely with the objective evaluation result. Projected StyleGAN2 (VGG$+$SCNN$_\text{Loop}$) performs the best and Projected StyleGAN2 (SCNN$_\text{Loop}$) performs the worst. 
Overall, Projected StyleGAN2 (VGG$+$SCNN$_\text{Loop}$) achieves  results comparable to StyleGAN2 and even `Real' for both types of loops. 
Somehow surprisingly, Projected StyleGAN2 (VGG$+$SCNN$_\text{Loop}$) can even outperform `Real' in audio quality and  preference for both drum and synth loops. 
We conjecture that this is because the `Real' here actually stands for the ``MelGAN-vocoded'' version of the real data, whose quality may have suffered from the artefacts introduced by the vocoder.

Almost all models lead to higher subjective scores for drum loops than for synth loops, suggesting that  synth loop generation is more challenging.  
We note that, for synth loops, Projected StyleGAN2 (VGG$+$SCNN$_\text{Loop}$) actually obtains higher harmony scores than StyleGAN2 and `Real', but its loopability scores is lower. 
To further improve its performance for synth loop generation, future work may be done to explicitly consider loopability  in model training.


\section{Limitations}

We only focus on  one-bar loop generation with a specific BPM of 120 in our study, which is ``convenient'' as it gives us fixed-size Mel-spectrograms to be treated as images by StyleGAN2. However, to be applicable to loop-based music production, future work needs to consider variable BPMs and loops with more bars. 
As the size of the Mel-spectrogram would not be fixed when we consider variable BPMs, future work may need to adopt  generative models other than StyleGAN2 as the backbone. 
Possible candidates are UNAGAN \cite{unagan} and VQGAN \cite{vqgan}, both of which may also benefit from the ideas of Projected GAN.

To apply Projected GAN, we need to take care of the \emph{shape matching} between the pre-trained networks and the generative model. 
For example, in our work the pre-trained networks (e.g., the VGGish one) is trained on 1-second audio chunks, but our generator is to generate 2-second loops.  
The size mismatch can be easily addressed by simply splitting the Mel-spectrogram into two chunks in our work, but this is trickier if we consider BPMs other than 120.



\section{Conclusion}

In this paper, we have demonstrated that Projected GAN can be used to improve the training efficiency and overall performance of GAN-based models for audio generation, specifically the generation of drum loops and synth loops.
Moreover, we demonstrated that using a domain-specific pre-trained feature network alone does not work well; we need to use either a general pre-trained feature network, or the fusion of multiple pre-trained feature networks.


This work can be extended in many ways. First, we can expand our work to generate variable-length loops, or to other GAN-based audio-related tasks. 
Next, we can explore unsupervised or self-supervised approaches (e.g., \cite{DBLP:journals/corr/abs-2011-10566,DBLP:journals/corr/abs-2111-06377}) to get the pre-trained feature network. 
Third, we are also interested in using class conditions or attributes for more controllable loop generation.
Doing so may help improve the diversity of the generated loops as well, according to related work on image generation \cite{styleganxl}. Finally, using diffusion probabilistic models \cite{diffgantts,diffusiongan} as the generative model for loop generation also worth trying.

\section{Acknowledgement}
We are grateful to Tun-Min Hung for helpful discussions during the project. We also thank the anonymous reviewers for their constructive feedbacks. Our research is funded by grant NSTC 109-2628-E-001-002-MY2 from the National Science and Technology Council of Taiwan.

\bibliography{ISMIRtemplate}

\end{document}